\documentclass[aps,twocolumn,showpacs,superscriptaddress]{revtex4}
\usepackage[dvips]{graphics,graphicx}

\begin{document}
\title{The driven Harper model}
\author{Andrey R. Kolovsky}
\affiliation{Kirensky Institute of Physics, 660036 Krasnoyarsk, Russia}
\affiliation{Siberian Federal University, 660041 Krasnoyarsk, Russia}

\author{Giorgio Mantica}
\affiliation{Center for Nonlinear and Complex Systems,
University of Insubria, 22100 Como, Italy}
\affiliation{CNISM unit\`a di Como}  \affiliation{I.N.F.N. sezione di Milano, Italy}
\date{\today}

\begin{abstract}
We analyze the driven Harper model, which appears in the problem of tight-binding electrons in the Hall configuration (normal to the lattice plane magnetic field plus in-plane electric field). The presence of an electric field extends the celebrated Harper model, which is parametrized by the Peierls phase, into the driven Harper model, which is additionally parametrized by two Bloch frequencies associated with the two components of the electric field. We show that the eigenstates of the driven Harper model are either extended or localized, depending on the commensurability of Bloch frequencies. This results holds for both rational and irrational values of the Peierls phase. In the case of incommensurate Bloch frequencies we provide an estimate for the wave-function localization length.
\end{abstract}
\maketitle

{\em 1.} The Harper Hamiltonian naturally appears in the problem of crystal electrons in the presence of a magnetic field \cite{Harp55}. It describes the energy spectrum of a tight-binding electron in a 2D lattice, whose graphic representation is widely known as the Hofstadter butterfly \cite{Hofs76}. The Harper Hamiltonian is a particular case, $J_x=J_y$, of the more general Aubry-Andr\'e model \cite{Aubr80},
\begin{equation}
\label{1}
 (\widehat{H} b)_l = -\frac{J_x}{2}\left(b_{l+1} +  b_{l-1}\right) - J_y\cos(2\pi\alpha l)b_l \;,
\end{equation}
where the notations explicitly refer to tight-binding electrons in a 2D square lattice: $J_x$ and $J_y$ are the hopping matrix elements along the primary axes and $\alpha$ is the Peierls phase, given by the magnetic flux through the elementary cell. If the parameter $\alpha$ in (\ref{1}) is an irrational number, the spectrum of $\widehat{H}$ is known to be pure point for $J_y>J_x$, continuous for  $J_y<J_x $, and singular continuous for $J_y=J_x$. Because of this remarkable feature the system (\ref{1}) also serves as a model of Anderson localization in quasicrystals \cite{Aubr80,Roat08}.

In this letter we discuss the driven Harper model, for which the  Schr\"odinger equation for the time-dependent quantum amplitudes $b_l$ reads
\begin{equation}
\label{2}
i\dot{b}_l=-\frac{J_x}{2}\left(e^{-i\omega_xt} b_{l+1} + e^{i\omega_xt} b_{l-1}\right)
-J_y\cos(2\pi\alpha l + \omega_y t) \; b_l \;.
\end{equation}
This model was introduced in our recent publications \cite{85,preprint,86} devoted to cold atoms in a 2D optical lattice, subject to an artificial magnetic field normal to the lattice plane and to an in-plane static (for example, gravitational) force. Clearly, this model also describes a tight-binding electron in the Hall configuration, so that our results can be equally applied to this fundamental solid-state system. In this case, the frequencies $\omega_x$ and $\omega_y$ are the Bloch frequencies associated with the components of the electric field.

The energy spectrum and the eigenstates of an electron in the Hall configuration crucially depend on the commensurability of Bloch frequencies. Namely, for any rational ratio $\omega_x/\omega_y=r/q$ (here $r,q$ are co-prime numbers) the energy spectrum is continuous and the eigenstates are extended functions \cite{preprint,Naka95}. It was  conjectured in Ref.~\cite{preprint} that for incommensurate  Bloch frequencies the energy spectrum is discrete, but the localized eigenfunctions are characterized by a non-polynomial scaling law, with deep implications on the time dynamics of the system. The analysis of the 1D system (\ref{2}) presented in this paper identifies this scaling law, proves the discrete nature of the spectrum for irrational $\beta=\omega_x/\omega_y$ and explains the interesting dynamical phenomena described in \cite{preprint}.

{\em 2.} We begin with a semiclassical analysis of the driven Harper model. The classical counterpart of (\ref{2}) reads
\begin{equation}
\label{3}
H_{cl}(t)=-J'_x\cos(p-\omega_x t) - J'_y\cos(x+\omega_y t)  \;,
\end{equation}
where $p$ and $x$ are canonically conjugated variables and $J'_{x,y}=2\pi\alpha J_{x,y}$. In fact, it can be shown that Eq.~(\ref{2}) follows from (\ref{3}) if $p$ and $x$ are operators obeying the commutation relation $[\hat{x},\hat{p}]=i2\pi\alpha$, so that the Peierls phase $\alpha$ plays the role of an effective Planck constant.  The system (\ref{3}) can be equally studied on the torus ($-\pi\le p,x <\pi$), on the cylinder  ($-\pi\le p <\pi$, $-\infty<x<\infty$), and in the plane ($-\infty<p,x<\infty$). Considering the last case, it is easy to prove that the system (\ref{3}) is completely integrable. Indeed, using the canonical substitution $p'=p-\omega_x t$ and $x'=x+\omega_y t$ the new Hamiltonian appears to be time-independent,
\begin{equation}
\label{4}
H'_{cl}=-J'_x\cos(p') - J'_y\cos(x') +\omega_x x'+\omega_y p'  \;,
\end{equation}
and, hence, the right hand side of (\ref{4}) is the global integral of the motion. In what follows, however, we shall discuss the dynamics of the classical system (\ref{3}) on the cylinder, that can be compared with the quantum system (\ref{2}).
We are interested in the particle motion along the $x$ axis, and in particular, in its mean velocity $\bar{v}=\lim_{t\rightarrow\infty} x(t)/t$.

Although the system (\ref{3}) is completely integrable for any set of parameters, it has qualitatively different dynamical regimes, depending on whether $\beta=\omega_x/\omega_y$ is a rational number, and depending on the relative value of $\Omega$,
the system characteristic frequency in the absence of driving,
\begin{equation}
\label{5}
\Omega=(J'_x J'_y)^{1/2}=2\pi\alpha  (J_x J_y)^{1/2} \;,
\end{equation}
versus $\omega$, the geometric sum of the driving frequencies $\omega=\sqrt{\omega_x^2+\omega_y^2}$ \cite{remark1}.  In the high-frequency regime, $\omega\gg\Omega$, and for irrational $\beta$ any phase trajectory is bounded, implying $\bar{v}=0$. The particle can have nonzero mean velocity only if $\beta$ is a rational number. This can be proved using adiabatic perturbation theory, where one distinguishes between the fast variables $p',x'$ and the slow variables $p,x$. We demonstrate this for two particular cases: $\beta=0$ and $\beta=1$.
If $\beta=0$ the slow variable $p(t)\approx p_0$ where $p_0$ is the initial momentum. Then $x\approx x_0+J'_x\sin(p_0) t$, and $\bar{v}=J'_x\sin(p_0)$. If a classical ensemble of particles is uniformly distributed over the `elementary cell' $-\pi\le p,x <\pi$, we obviously obtain ballistic spreading, where the mean-squared displacement  $\sigma=\sqrt{\langle x^2 \rangle -\langle x \rangle^2}$ asymptotically follows a linear law, $\sigma(t)=At$, with $A=J'_x/\sqrt{2}$. Next, consider the case $\beta=1$. As in the former, at zero order we have $p(t)=p'+\omega_x t=p_0$. However, at first  order, the momentum $p(t)=p_0+(J'_y/\omega_y)\cos(x+\omega_y t)$ is a periodic function of time.  Substituting this solution into the Hamiltonian equation for the conjugate variable $x$ we have $x(t)=J'_x\int_0^t \sin[p(t)-\omega_x t] {\rm d}t \sim t$, where the proportionality coefficient can be expressed  through the Bessel function ${\cal J}_1(J'_y/\omega_y)$.  This implies that the ensemble of particles has a dispersion $\sigma(t)=At$ with $A\sim J'_x J'_y/\omega$. These rates of ballistic spreading are two particular cases of a general result,
\begin{equation}
\label{6}
A\sim\omega^{-(r+q-1)} \;,\quad \omega\gg \Omega \;,
\end{equation}
which coincides with the rate of wave-packet spreading derived in Ref.~\cite{preprint} by using quantum perturbation theory.  
\begin{figure}
\center
\includegraphics[width=8.5cm, clip]{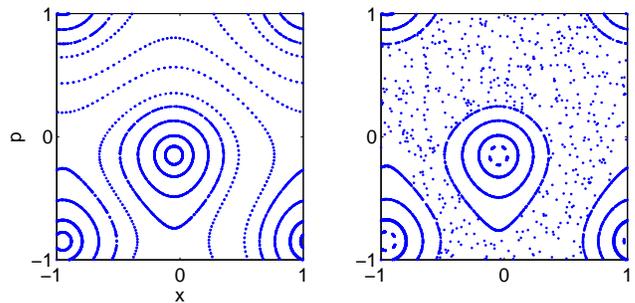}
\caption{A portion of phase space (stroboscopic map over $T_y=2\pi/\omega_y$) of the classical driven Harper model  for rational $\beta=1/3$, left panel, and irrational $\beta=(\sqrt{5}-1)/4\approx 1/3$, right panel. The other parameters are $J'_{x,y}=2\pi \cdot 0.1545$ and $\omega=0.3$. Transporting islands are seen as stability islands surrounding elliptic points at $(x,p)\approx(0,0)$ and $(x,p)\approx(-\pi,-\pi)$. In the case of rational $\beta$ phase trajectories are closed on the torus and the stroboscopic map reproduces these trajectories.  For irrational $\beta$ any trajectory, which does not belong a stability island, never repeats itself on the torus and appears as a scattered array of points resembling a chaotic trajectory.}
\label{fig1}
\end{figure}

\begin{figure}
\center
\includegraphics[width=8.5cm, clip]{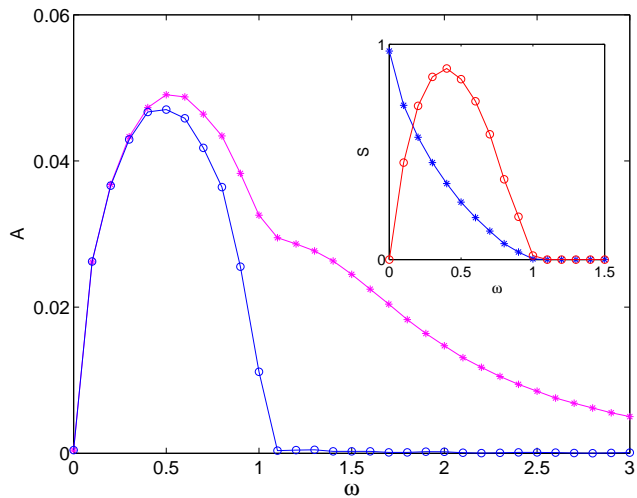}
\caption{The rate of ballistic spreading for an ensemble of classical particles, versus the driving frequency $\omega$, for rational $\beta=1/3$ (upper curve, stars) and irrational $\beta=(\sqrt{5}-1)/4$ (lower curve, open circles). The inset shows the relative size, $S$, of transporting islands (stars) for $\beta=(\sqrt{5}-1)/4$ and the function (\ref{6b}) (open circles), where we arbitrary set the proportionality coefficient to 4.}
\label{fig2}
\end{figure}

The regime of low-frequency driving, $\omega<\Omega$, is more subtle, because here the phase space of (\ref{3}) contains two chains of transporting islands, see Fig.~\ref{fig1}. Remark that these chains exist for both rational and irrational values of $\beta$. In a classical ensemble of particles, those with initial conditions in the transporting islands move in the negative direction at velocity $\bar{v}=\omega_y$, while the others travel in the positive direction, and $\sigma(t)=At$; the  values of the coefficient $A$ obtained numerically are depicted in Fig.~\ref{fig2} for the two values of $\beta$ used in Fig.~\ref{fig1}. As expected, when $\beta$ is a rational number, the linear dependence $A(\omega)\sim \omega$,  valid for $\omega\ll\Omega$, leaves place for large $\omega$ to the asymptotic dependence (\ref{6}). For irrational $\beta$ the rate of ballistic spreading  is given by
\begin{equation}
\label{6b}
A\sim\omega S(\omega) \;,
\end{equation}
where $S(\omega)$ is the relative size of transporting islands ($1\le S \le 0$, see inset in Fig.~\ref{fig2}).

\begin{figure}
\center
\includegraphics[width=8.5cm, clip]{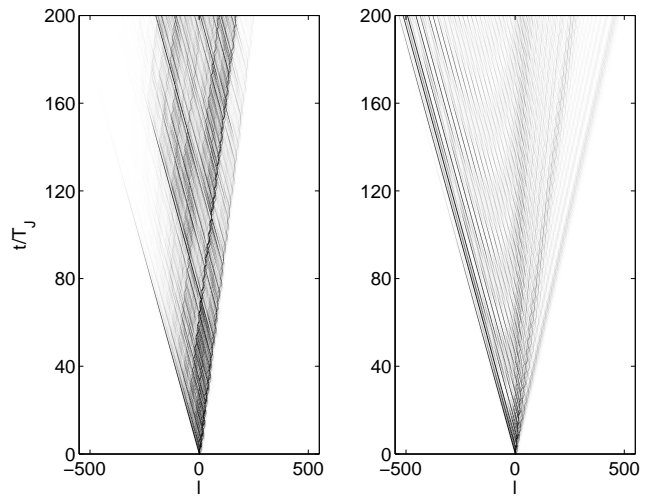}
\caption{Grey tone image of the quantum wave-packet (black maximum) versus space $l$ and time $t$.  The left panel is for irrational $\beta=(\sqrt{5}-1)/4\approx 1/3$, the right panel has $\beta=1/3$.
The other parameters are $J_x=J_y=1$, $\alpha=0.1545$ and $\omega=0.45$.}
\label{fig3}
\end{figure}

{\em 3.} We now turn to quantum analysis. An important feature of our system is the presence, for $\omega<\Omega$, of chains of transporting islands. A system of this kind -- the asymmetric kicked Harper -- was studied in a series of works by Ketzmerick {\em et al.} \cite{Hufn02,Back05}, showing that its quantum dynamics may considerably differ from the classical. Here we meet a similar situation, although our system has no chaotic component, at difference with that studied in \cite{Hufn02,Back05}. We have found that, even if the classical driven Harper shows a ballistic regime for both rational and irrational $\beta$, the quantum motion of the driven Harper is ballistic only for rational $\beta$, while for irrational $\beta$ a saturation effect takes place. In other words, for any irrational $\beta$, the wave-packet dispersion $\sigma(t)$ follows the linear law of the classical system only for finite times, being asymptotically bounded from above. The physics behind this phenomena is the destructive interference between two probability flows going in  opposite directions. Figure \ref{fig3} shows the evolution of a localized wave-packet, which is initially supported by the central transporting island. Tunneling out of this island as well as the opposite process of capture into other islands in the chain are clearly seen. The rate of tunneling is defined by the ratio between the size of the stability island, $S=S(\omega)$, and the effective Planck constant $\hbar_{eff}=2\pi\alpha$.  Following analog arguments in Ref.~\cite{Hufn02}, we can estimate the maximal wave-packet dispersion $\sigma_{max}$ as follows. Consider an initially populated transporting island. Due to tunneling it is depleted after a time which is exponential in $S/\hbar_{eff}$ \cite{remark2}. During this time the quantum particle is transported at distance $\omega_y/2\pi \alpha$ in units of the lattice periods. Therefore,
\begin{equation}
\label{7}
\sigma_{max}\sim \frac{\omega}{\alpha}\exp\left(C\frac{S(\omega)}{\alpha}\right) \;,
\end{equation}
where $C$ is some constant. It should be mentioned that tunneling in and out transporting islands also takes place in the case of rational $\beta$. However, for rational $\beta$ the interference between two probability flows is constructive and $\sigma(t)$ obeys a linear law for all times.

\begin{figure}[t!]
\center
\includegraphics[width=8.5cm, clip]{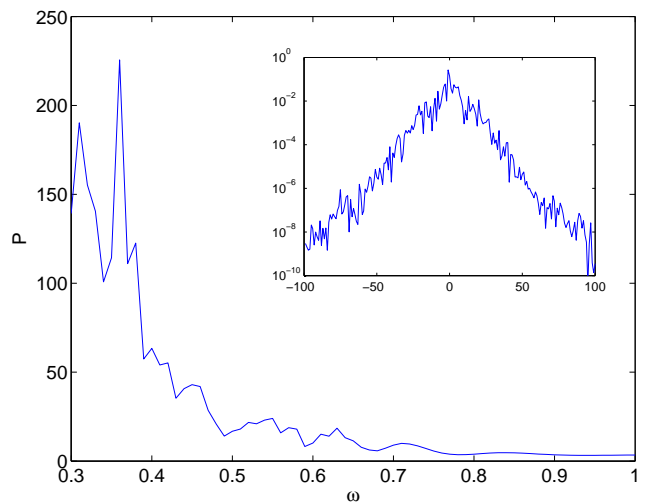}
\caption{The localization length of eigenstates of the evolution operator for $\beta=(\sqrt{5}-1)/4$. The mean participation ratio is shown as the function of $\omega$.
The inset shows a typical eigenstate for $\omega=0.6$.}
\label{fig4}
\end{figure}

The observed difference in the wave-packet dynamics for rational and irrational $\beta$ indicates a difference in spectral properties of the evolution operator $\widehat{U}$. We construct this operator by using the substitution $b_l \rightarrow b_l\exp(i\omega_x l t)$, which modifies Eq.~(\ref{2}) as follows
\begin{equation}
\label{8}
i\dot{b}_l=-\frac{J_x}{2}\left(b_{l+1} + b_{l-1}\right)
-J_y\cos(2\pi\alpha l + \omega_y t)b_l + \omega_x l b_l \;,
\end{equation}
and by integrating (\ref{8}) over the period $T_y=2\pi/\omega_y$. The obtained evolution operator can be approximated to arbitrary precision by a banded matrix, whose bandwidth depends on the parameter $J_x$. It is instructive to consider the case $J_x=0$, where the operator $\widehat{U}$ is a diagonal matrix with the elements $U_{l,l}=\exp(-i2\pi \beta l)$, which is a periodic (aperiodic) function of $l$ for rational (irrational) $\beta$.  [Notice that because of integration over time the parameter $\alpha$ does note appear in the last expression. This explains or, at least, gives a hint why the driven Harper model is insensitive  to rationality of the parameter $\alpha$.]  At this point we can draw an analogy with a paradigmatic model of quantum chaos -- the kicked rotor \cite{Casa79}. For vanishing kick amplitude, the evolution operator of the kicked rotor is also a diagonal matrix with matrix elements $U_{l,l}=\exp(-i2\pi \xi l^2)$, which is a periodic or aperiodic function of $l$ according to rationality of the parameter $\xi$. The number theoretic characteristics of $\xi$ are crucial \cite{italogiu}, since they determine whether the eigenfunctions are extended or localized. The driven Harper model shows the same feature, since the eigenfunctions of the evolution operator are localized functions of $l$ if the parameter $\beta$ is an irrational number (see inset in Fig.~\ref{fig4}).

We studied numerically the localization length as a function of the parameter $\omega$. Figure \ref{fig4} shows the participation ratio $P=1/\sum_l |b_l|^4$ averaged over 300 eigenstates. One clearly sees an exponential increase in the localization length with decrease of the driving frequency $\omega$, i.e., with increase of the size of transporting islands.  A similar exponential dependence  is also obtained when we vary $\alpha$ at fixed $\omega$ \cite{remark3}. These results confirm the estimate (\ref{7}), which can be equally used for  the maximal dispersion (saturation level) and the mean localization length.

{\em 4.} Conclusions. In the physical problem of tight-binding electrons in the Hall configuration, the transporting states, which are associated with classical islands, are responsible for quantum transport of electrons in a direction perpendicular to the vector of the electric field, i.e., for the Hall current \cite{85,87}. It was numerically observed in Ref.~\cite{preprint} that for irrational directions of the electric field this transport abruptly diminishes when the magnetic field (which in the tight-binding approximation defines the Peierls phase $\alpha$) is increased. At the same time, no sign of inhibited transport was observed for a small $\alpha$.  These features of the original system find natural explanation in terms of the 1D model (\ref{2}), where the localization length grows exponentially with the inverse of the parameter $\alpha$.  Moreover, the driven Harper model studied here is interesting in its own right, since it can be realized in laboratory experiments with cold atoms in quasi 1D optical lattices \cite{86}. Detailed studies of this application, together with the driven Harper dynamics, performed in line with Ref.~\cite{Hufn02,Back05}, will be presented elsewhere.

Acknowledgements. Computations for this work have been performed on the CSN4 cluster of INFN in Pisa. G.M. acknowledges the support of MIUR-PRIN project {\em Nonlinearity and disorder in classical and quantum transport processes} and A.K.  acknowledges the support of SB RAS project {\em Dynamics of atomic Bose-Einstein condensates in optical lattices} and RFBR project {\em Tunneling of the macroscopic quantum states}.


\begin{thebibliography}{10}
\bibitem{remark1}
In the original physical problem $\omega$ is proportional to the magnitude of the electric field, while $\Omega$ is the cyclotron frequency, proportional to the magnetic field magnitude.
\bibitem{remark2}
Numerical studies of the tunneling rate indicate the presence of a resonance-like structure above the overall exponential dependence. In particular, there is a local deep in the tunneling rate at $\omega\approx 0.36$ which is responsible for a local peak in $\sigma_{max}=\sigma_{max}(\omega)$.
\bibitem{remark3}
We note that to study the quantum-classical correspondence with respect to $\hbar_{eff}=2\pi\alpha$ one fixes the scaled parameters $J'_{x,y}=2\pi\alpha J_{x,y}$, which leaves the classical dynamics unchanged.
\end{thebibliography}
\end{document}